\documentclass[journal=cmatex,layout=draft]{achemso}
\setcitestyle{square}
\usepackage[version=4]{mhchem}
\usepackage[hyperindex]{hyperref}
\hypersetup{colorlinks,citecolor=blue,linkcolor=blue,urlcolor=blue}
\usepackage{units}
\usepackage{amsmath}
\usepackage{booktabs}
\usepackage{etoolbox}

\newcommand{\diffd}{\,\text{d}}
\newcommand{\jph}{j_\text{ph}}
\newcommand{\dcrit}{d_\text{crit}}

\newcommand{\LD}{\lambda_\text{D}}

\makeatletter
\patchcmd{\acs@contact@details}{E}{*\,E}{}{}
\makeatother

\author{Sebastian Wilken}
\affiliation{Institute of Physics, Carl von Ossietzky University of Oldenburg, 26111 Oldenburg, Germany}
\alsoaffiliation{Physics, \AA{}bo Akademi University, Porthansgatan 3, 20500 Turku, Finland}
\email{sebastian.wilken@uol.de}

\author{Oskar J. Sandberg}
\affiliation{Department of Physics, Swansea University, Singleton Park, Swansea, SA2\,8PP, Wales, UK}

\author{Dorothea Scheunemann}
\affiliation{Institute of Physics, Carl von Ossietzky University of Oldenburg, 26111 Oldenburg, Germany}
\alsoaffiliation{Physics, \AA{}bo Akademi University, Porthansgatan 3, 20500 Turku, Finland}

\author{Ronald \"{O}sterbacka}
\affiliation{Physics, \AA{}bo Akademi University, Porthansgatan 3, 20500 Turku, Finland}

\title{Watching Space Charge Build up in an Organic Solar Cell\footnote{This publication is dedicated to Prof.~J\"urgen Parisi on the occasion of his retirement}}

\SectionNumbersOn

\begin{document}

\begin{abstract}
Space charge effects can significantly degrade charge collection in organic photovoltaics (OPVs), especially in thick-film devices. The two main causes of space charge are doping and imbalanced transport. Although these are completely different phenomena, they lead to the same voltage dependence of the photocurrent, making them difficult to distinguish. In this work, a method is introduced how the build-up of space charge due to imbalanced transport can be monitored in a real operating organic solar cell. The method is based on the reconstruction of quantum efficiency spectra and requires only optical input parameters that are straightforward to measure. This makes it suitable for the screening of new OPV materials. Furthermore, numerical and analytical means are derived to predict the impact of imbalanced transport on the charge collection. It is shown that when charge recombination is sufficiently reduced, balanced transport is not a necessary condition for efficient thick-film OPVs.
\end{abstract}

\section{Introduction}
The efficiency of organic photovoltaics~(OPVs) has increased from about 10 to over 16\% within only a few years.\cite{Cui2019,Hong2019} The main driver of this rapid progress is emergence of non-fullerene acceptors. These materials have opened a completely new parameter space for researchers to find efficient donor--acceptor combinations. However, still very few systems are known that maintain their full performance at technically relevant thicknesses of~\unit[300]{nm} and more.\cite{Armin2015,Jin2017,Sun2018,Zhang2018} The reason is that the operation of modern OPVs resembles a competition between charge collection and charge recombination driven by the internal electric field.\cite{Bartesaghi2015,Neher2016,Heiber2018} Increasing the thickness inevitably slows down collection, which in turn increases the probability for photogenerated carriers to recombine before reaching the electrodes. Since the mobility is limited by the hopping nature of transport~(typically below $\unit[10^{-2}]{cm^2\,V^{-1}\,s^{-1}}$), reducing the recombination has been identified as key strategy to realize thick-film devices with high fill factor and quantum efficiency.\cite{Mauer2010,Bartelt2015,Gohler2018}

This simple picture becomes complicated when space charge is present.\cite{Stolterfoht2016,Kaienburg2016} The effect of space charge is to redistribute the electric field in the active layer into two parts, a space charge region~(SCR) and a quasi-neutral region. In the quasi-neutral region, the internal field is screened, and charge transport is carried out by diffusion only. This makes carriers much more sensitive to recombination and typically leads to a decreased device performance. Conversely, the internal field is enhanced within the SCR, which may even improve collection from this part of the active layer. Hence, the photocurrent~($\jph$) will be dominated by carriers generated in the SCR and critically depends on the SCR width. Space-charge effects in OPVs may occur due to (unintentional) doping, either within the active layer or in the vicinity of the contacts.\cite{Dibb2013,Deledalle2015,Schafferhans2010,Nyman2016,Sandberg2019} Another source of space charge is imbalanced charge transport.\cite{Scheunemann2019,Mihailetchi2005,Kaienburg2016,Kirchartz2012,Stolterfoht2016,Spies2017} If there is a significant mismatch between the mobility of electrons~($\mu_n$) and holes~($\mu_p$) in the active layer, charges will accumulate at the electrode extracting the slower carrier. For example, if holes are slower than electrons, the SCR will form at the anode, while the electric field is zero in the vicinity of the cathode.

Even though doping and imbalanced transport are two fundamentally different phenomena, they lead to the same electric-field dependence of the photocurrent: $\jph \propto V^{1/2}$, where $V$ is the applied voltage.\cite{Mihailetchi2005,Sandberg2019,Deledalle2015} Hence, from a simple current--voltage measurement, one cannot decide whether the degraded device performance is caused by doping or imbalanced transport. The key lies instead in the light-intensity dependence. In the doping case, the additional carrier population causing the space charge is fixed by the density of dopants. We have recently shown that this leads to a collection efficiency that is independent of the light intensity.\cite{Sandberg2019} In contrast, for imbalanced transport, the pile-up of slower carriers becomes more and more pronounced as the photogeneration increases. The width of the SCR consequently gets smaller, which let the photocurrent become sub-linear and scale as~$G^{3/4}$ with the generation rate~($G$).\cite{Mihailetchi2005,Goodman1971} So one might think that to prove that a device is limited by imbalanced transport, it is sufficient to measure the scaling of the photocurrent with light intensity, $\jph \propto I^\beta$, and check whether $\beta = 3/4$ is fulfilled. However, this approach is prone to error as changes in~$\beta$ are often subtle and hard to detect.\cite{Koster2011,Deibel2010,Dibb2011} It is also known that bimolecular recombination alone can result in an arbitrary exponent between $1$ and $1/2$, even if no space charge is present.\cite{Koster2011} Instead, it would be desirable to directly probe the width of the SCR and its change with light intensity.

In this paper, we present a simple and robust method how the build-up of space charge due to imbalanced transport can be monitored in a real operating device. The method is based on the reconstruction of white-light bias external quantum efficiency~(EQE) spectra and is thoroughly tested against numerical simulations. We demonstrate our approach for an experimental system with a mobility mismatch of one order of magnitude. We show that charge collection becomes a function of the position and the collection zone varies according to an analytical model for imbalanced transport. Importantly, our approach does not require knowledge about the electron and hole mobility or other electrical parameters like the recombination rate constant. This makes it valuable for the screening of new blend systems for efficient thick-film OPVs that are compatible with large-area manufacturing methods. We finally use our device model to give a conclusive view on the thickness limits for OPVs with imbalanced transport. 

\section{Results and Discussion}
\subsection{Analytical Model}
We begin with an analytical model for a device limited by imbalanced transport. As shown by Goodman and Rose,\cite{Goodman1971} and later adapted to OPVs by Mihailetchi~et~al.,\cite{Mihailetchi2005} the photocurrent in the SCR will be space-charge-limited, taking the form $\jph = -qG^{3/4}(9\mu_s\varepsilon\varepsilon_0/8q)^{1/4}(V_0 - V)^{1/2}$, where $\mu_s$ is the mobility of the slower carrier, $\varepsilon\varepsilon_0$ is the permittivity of the active layer, $q$ the elementary charge, and $V_0$ the potential drop across the SCR at $V = 0$. This expression, however, does not account for the spatial dependence of the electron and hole currents in the SCR, where recombination is negligible. In the Supporting Information we establish that if these currents are correctly considered, the photocurrent modifies to
\begin{equation}
\jph = -q G^{3/4} \left(\frac{4\mu_s\varepsilon\varepsilon_0}{q}\right)^{1/4} (V_0-V)^{1/2}.
\label{eq:jph}
\end{equation}
Note that Equation~(\ref{eq:jph}) shows the same scaling behavior with voltage and generation rate as reported previously, but differs by $\sqrt[4]{32/9} \approx 1.37$ from the Goodman and Rose result. The corresponding width~$w$ of the SCR is given by
\begin{equation}
w = \sqrt{2(V_0-V)}\left(\frac{\varepsilon\varepsilon_0\mu_s}{qG}\right)^{1/4}.
\label{eq:SCR}
\end{equation}
Our central hypothesis is that only an experimental test of Equation~(\ref{eq:SCR}) can unambiguously prove that a device is limited by imbalanced transport. In particular, it must be shown that the SCR width scales with a $-1/4$ power dependence on light intensity.

\subsection{Position-Dependent Charge Collection}
\begin{figure*}[t]
\includegraphics[width=\linewidth]{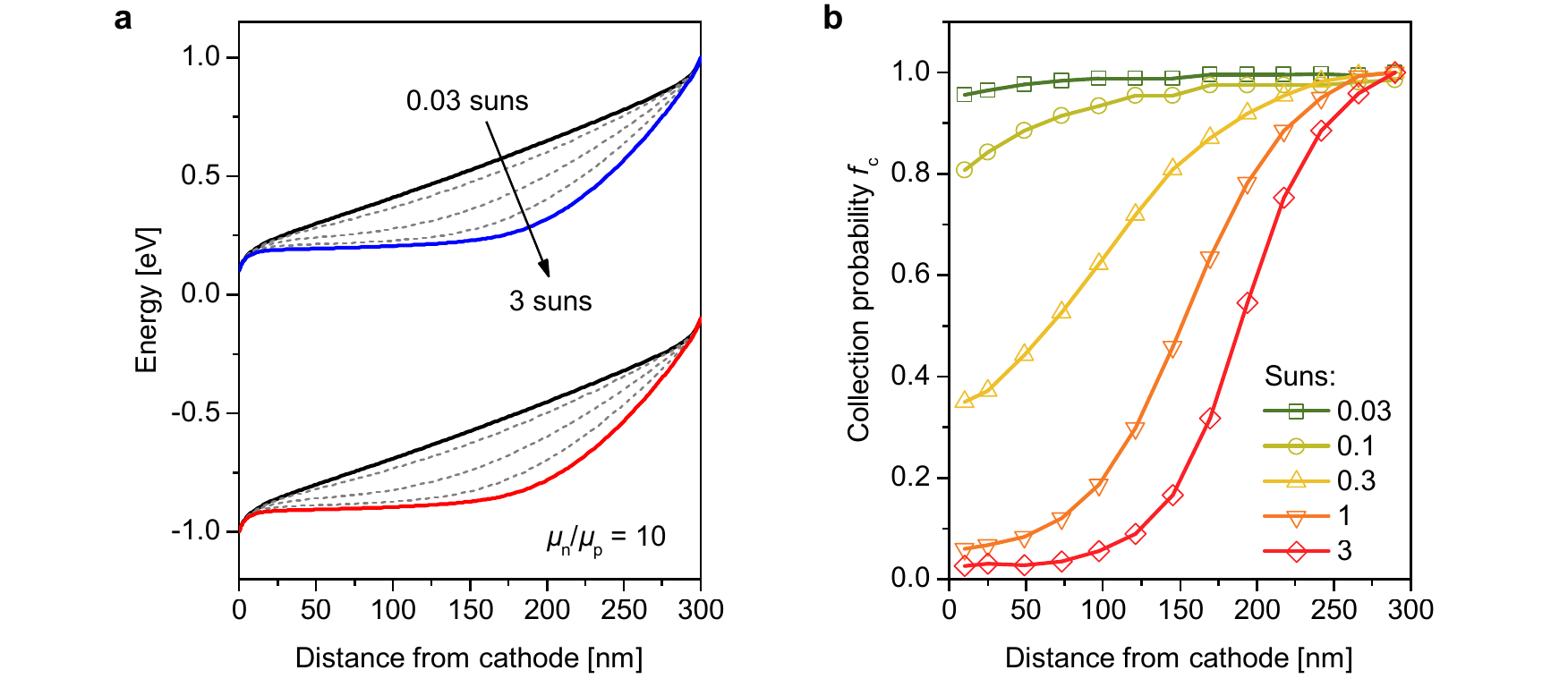}
\caption{The build-up of space charge due to imbalanced transport in a thick-film organic solar cell. (a)~Simulated energy level diagram under short-circuit conditions for a 300-nm device with a mobility mismatch of one order of magnitude~($\mu_n = 10 \mu_p = \unit[10^{-3}]{cm^2\,V^{-1}\,s^{-1}}$) at different light intensities. (b)~Corresponding charge-collection probability as a function of the position in the active layer. All illumination levels are given relative to air mass~1.5 spectral irradiance at $\unit[100]{mW\,cm^{-2}}$~(`\unit[1]{sun}').}
\label{fig:figure01}
\end{figure*}

We have recently used a numerical drift--diffusion model to discuss how imbalanced charge transport affects the open-circuit voltage.\cite{Scheunemann2019} Here, we will apply the same model to study the charge collection under short-circuit conditions. Figure~\ref{fig:figure01}a illustrates the effect of space charge on the simulated band diagram for a 300-nm thick organic solar cell with a mobility mismatch of one order of magnitude~($\mu_n/\mu_p = 10$). At low light intensity~(\unit[0.03]{suns}), the electric field is homogeneous over the active layer, resulting in a linear gradient of the energy bands. The situation changes drastically as the light intensity is increased: The pile-up of the slower holes leads to the formation of a SCR at the anode side, where the entire band bending is concentrated. It can be clearly seen that the width~$w$ of the SCR decreases with increasing generation rate, thereby increasing the width~$d-w$ of the region where the electric field is zero. Given the limited diffusion length in organic semiconductors, it is intuitively clear that most of the carriers photogenerated in the field-free region will undergo recombination instead of getting collected at the electrodes.

In order to quantify these losses, we consider the spatially resolved charge-collection probability~$f_c(x)$.\cite{Kirchartz2016,Kirchartz2012} The collection probability serves as a weighting factor for the generation rate to predict the photocurrent, $\jph = q \int f_c(x) G(x) \diffd x$. Note that due to the pronounced thin-film interference effects in OPVs, the generation rate is generally a non-trivial function of the position. In an ideal device, $f_c$ would be unity, so that the photocurrent would reach its maximum value~$\jph = q\bar{G}d$, where $\bar{G}$ denotes the spatial average of the generation rate. In our numerical model, we implement the collection probability by adding a small peak generation to the constant background generation rate and letting it shift from the back to the front of the active layer. The extra photocurrent due to the peak is then normalized to the case without background illumination, i.e., when the device is not disturbed by space charge effects. This way, the model mimics the white-light bias EQE measurements described below, but also other experiments in which a fixed carrier population is probed by applying a small perturbation, such as transient photocurrent and transient photovoltage. Figure~\ref{fig:figure01}b shows the resulting collection probability for the modeled device. Clearly, increasing the photogeneration induces a transition from homogeneous to inhomogeneous collection. Around 1-sun intensity, $f_c(x)$ is almost binary and changes abruptly from zero at the cathode to unity at the anode. This shape of~$f_c$ and, more importantly, its evolution with light intensity is unique for imbalanced transport and fundamentally different from the case without space charge or the doping case~(see the Supporting Information).

We point out that a direct observation of the potential distribution and the collection probability as illustrated in Figure~\ref{fig:figure01} would in principle be possible with scanning probe techniques such as Kelvin probe microscopy\cite{Weigel2015} and electron-beam induced currents.\cite{Abou-Ras2019,Ng2014} These techniques, however, are complex to carry out and usually not well suited to organic materials. In the following we will introduce a method that is much more simple and requires only standard equipment available in most OPV laboratories.

\subsection{Experimental Validation}
\begin{figure*}[t]
\includegraphics[width=\linewidth]{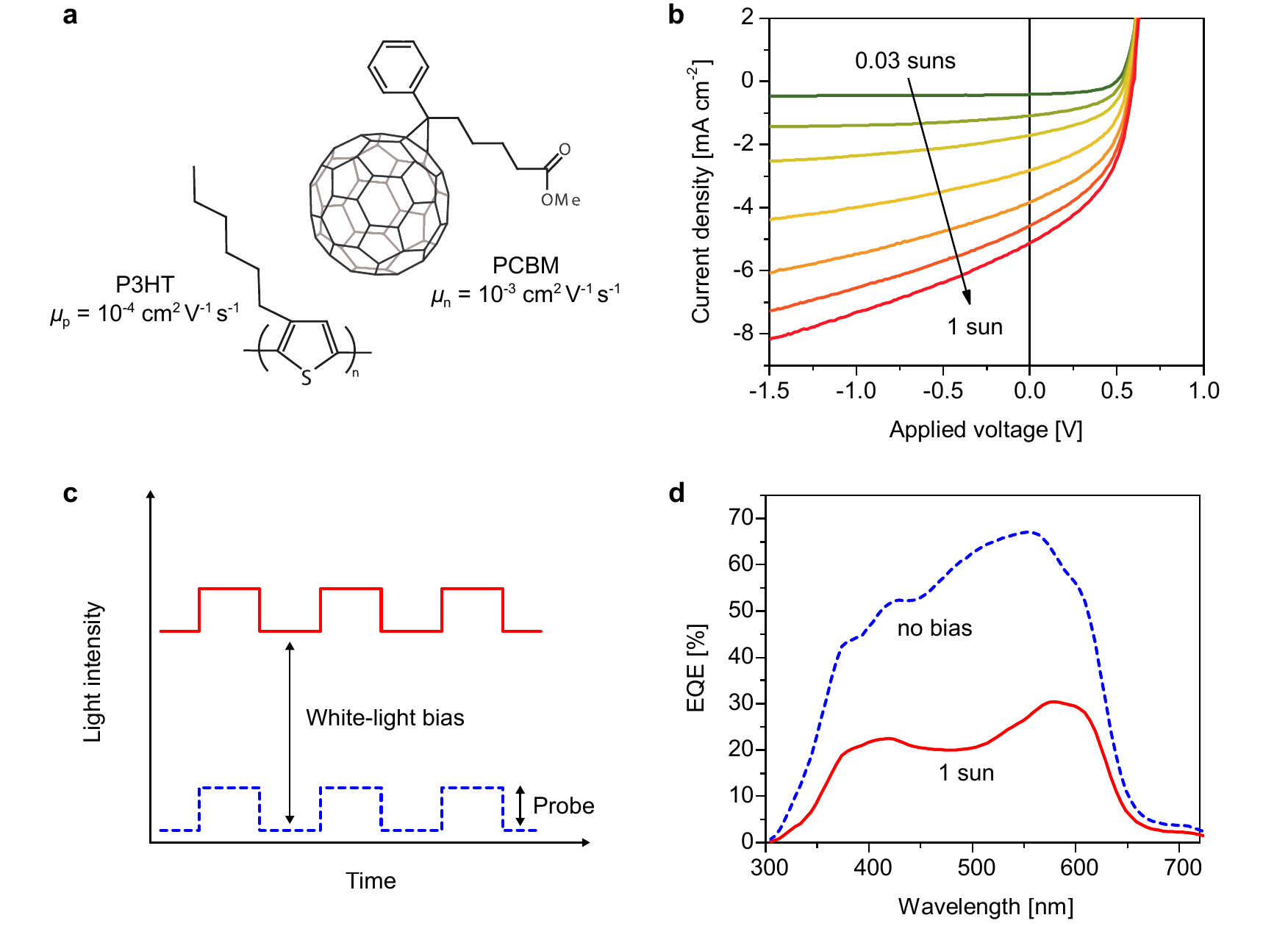}
\caption{Experimental system to demonstrate the effect of imbalanced transport. (a)~Structural formulas of P3HT and PCBM. (b)~Light-intensity dependent current--voltage curves for a 300-nm thick, inverted P3HT:PCBM solar cell. (c)~Principle of white-light bias EQE measurements. (d)~Low-intensity EQE spectra of the thick-film P3HT:PCBM device without~(dashed line) and with a white-light bias of \unit[1]{sun} intensity~(solid line).}
\label{fig:figure02}
\end{figure*}

To check if the predictions of the model are relevant to real devices, we used the well known blend of regioregular poly(3-hexylthiophene) and [6,6]-phenyl-\ce{C61}-butyric acid methyl ester~(P3HT:PCBM, Figure~\ref{fig:figure02}a). For the purpose of this study, P3HT:PCBM is an ideal test system. First of all, with an electron mobility of about~$\unit[10^{-3}]{cm^2\,V^{-1}\,s^{-1}}$ and a hole mobility of about~$\unit[10^{-4}]{cm^2\,V^{-1}\,s^{-1}}$, the mobility contrast is well documented.\cite{Bartelt2015,Kniepert2014,Shoaee2019} Second, because of the semicrystalline nature of P3HT:PCBM with relatively low energetic disorder, charge extraction and recombination are well described by quasi-equilibrium concepts, so that drift--diffusion simulations lead to meaningful results.\cite{Bartelt2015,Melianas2017} Third, the bimolecular recombination strength can be adjusted via the preparation conditions, in particular the drying rate of the solvent and the application of thermal annealing.\cite{Hamilton2010,Kniepert2014,Heiber2018} Here, we use blends that were rapidly dried and subsequently thermally annealed, which results in a bimolecular recombination rate constant of~$k_2 \approx \unit[10^{-12}]{cm^3\,s^{-1}}$~(see the Supporting Information). With this the recombination is two orders of magnitude reduced compared to the Langevin model, but still larger than for an optimized, solvent-annealed P3HT:PCBM~blend.\cite{Heiber2018}

Figure~\ref{fig:figure02}b shows light-intensity dependent current--voltage curves for a 300-nm thick P3HT:PCBM solar cell in the inverted architecture~(substrate\slash{}cathode\slash{}active layer\slash{}anode). Clearly, the photocurrent becomes voltage depend with increasing light intensity, which reduces the fill factor and the short-circuit current. Under standard operating conditions, the device reaches an open-circuit voltage of~\unit[0.59]{V}, a short-circuit current of~$\unit[5.1]{mA\,cm^{-2}}$, a fill factor of~0.41, and an efficiency of 1.2\%, which is well below what is expected for an optimized P3HT:PCBM solar cell.\cite{Dang2011} To learn more about the charge collection, we use white-light biased EQE measurements~(Figure~\ref{fig:figure02}c). In this technique, the current response to a small monochromatic probe is monitored versus different background photogeneration rates using lock-in technique. Hence, the quantity actually measured is the \textit{differential} spectral responsivity~(see Experimental Section), which is much more sensitive to intensity-dependent losses than the total photocurrent.\cite{Deibel2010,Dibb2011,Cowan2013,Wehenkel2012}. At low background illumination, the EQE peaks at around 70\%, which is indicative of efficient charge collection~(Figure~\ref{fig:figure02}d). The situation changes dramatically under 1-sun conditions. First, the overall height of the EQE is reduced to $\approx25\%$. Second, and most importantly, also the spectral shape of the EQE changes. It is obvious that the intensity-dependent losses are most pronounced at wavelengths~($\lambda$) around \unit[500]{nm}, i.e., the region in which P3HT:PCBM absorbs most strongly. In this spectral range, the generation profile is particularly inhomogeneous because the majority of photons are absorbed near the transparent cathode. Hence, it seems that the probability whether a charge carrier is collected depends on the position where it was generated in the active layer.\cite{Armin2015}

\begin{figure*}[t]
\includegraphics[width=\linewidth]{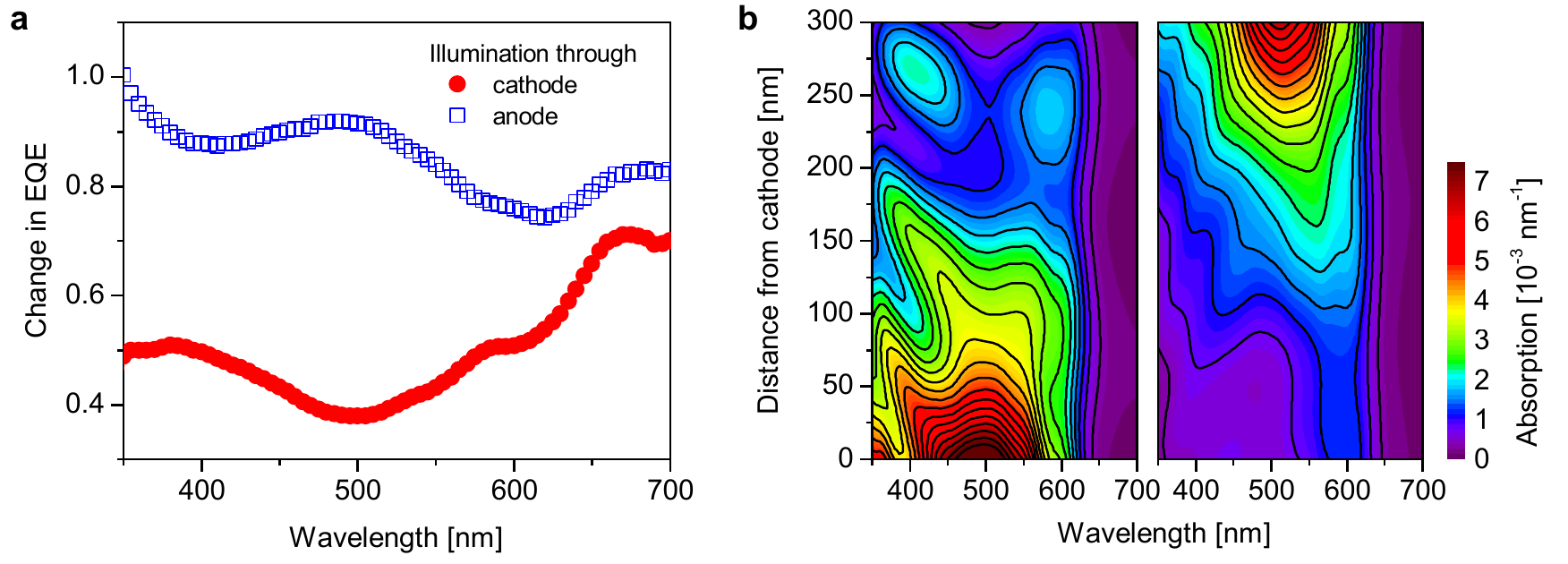}
\caption{Charge collection in semitransparent P3HT:PCBM devices. (a)~1-sun white-light bias EQE normalized to the case without background illumination for the device excited through the cathode and the anode, respectively. (b)~Corresponding absorption profiles~$A(x,\lambda)$ from transfer-matrix calculations. Left side shows the situation for illumination through the cathode, right side for illumination through the anode.}
\label{fig:figure03}
\end{figure*}

To prove that the collection becomes position-dependent, we fabricated semitransparent devices by exchanging the opaque hole-collecting electrode with a transparent insulator\slash{}metal\slash{}insulator structure based on ultra-thin Au films.\cite{Wilken2015} Using this approach, a strong variation of the generation profile can be obtained in the very same device by applying the illumination either through the bottom electrode~(here: the cathode) or the top electrode~(anode). Figure~\ref{fig:figure03} shows the change of the EQE at 1-sun white-light bias~(relative to the case with no background light) for the two illumination directions, and compares it with the spatially and spectrally resolved absorption profiles~$A(x,\lambda)$ from transfer-matrix calculations~(see Experimental Section). Two main features are clearly seen: (i)~collection losses are generally higher when the device is excited through the cathode rather than the anode; (ii)~the shape of the normalized EQE is nearly mirror-symmetric, with a minimum~(maximum) for illumination through the cathode~(anode) when the absorption profile is Beer--Lambert-like~($\lambda \approx \unit[500]{nm}$), and maxima~(minima) at wavelengths where it is more uniformly distributed throughout the active layer. Both findings confirm that the charge collection is position-dependent, with a relatively high collection probability in the vicinity of the anode and a relatively low collection probability in the vicinity of the cathode.

\subsection{Determination of SCR Width}
\begin{figure*}[t!]
\includegraphics[width=\linewidth]{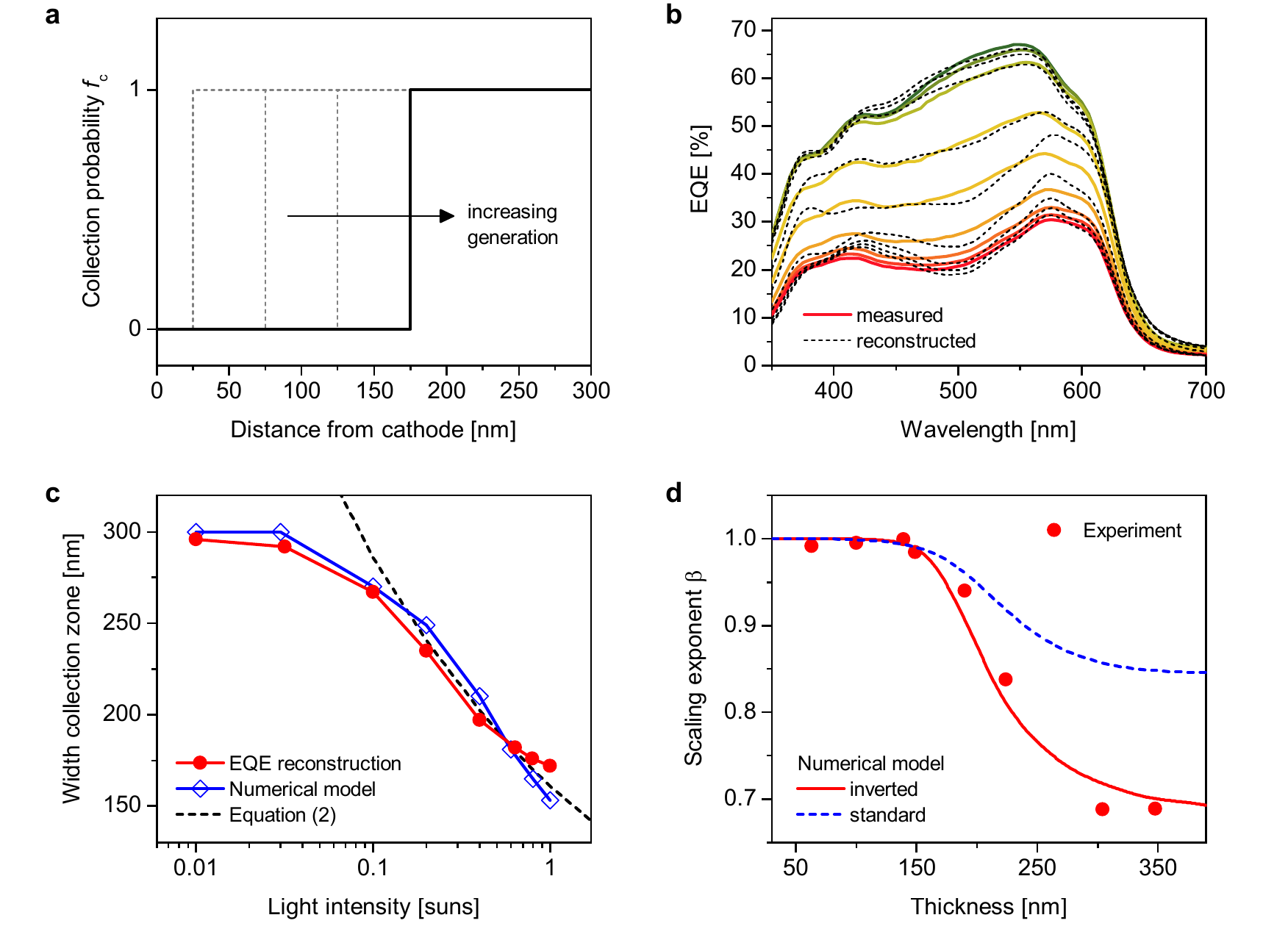}
\caption{Numerical reconstruction of EQE spectra. (a)~Illustration of the assumed collection probability~$f_c(x)$ for different photogeneration rates. (b)~Measured and reconstructed EQE spectra for a 300-nm thick P3HT:PCBM solar cell. (c)~Comparison of the derived width of the collection zone with the numerical device model (as effective collection width from the numerical model we took the point at which the collection probability~$f_c$ has dropped to half) and the analytical prediction from Equation~(\ref{eq:SCR}). (d)~Thickness dependence of the scaling exponent~$\beta$ of the photocurrent with respect to light intensity. Symbols are experimental data derived from the white-light bias EQE measurements as described in the Experimental Section. Lines are the prediction from the numerical model for an inverted~(substrate\slash{}cathode\slash{}active layer\slash{}anode) and a standard~(substrate\slash{}anode\slash{}active layer\slash{}cathode) device architecture.}
\label{fig:figure04}
\end{figure*}

Our goal now is to disentangle the spectral and spatial information from the EQE measurements also quantitatively. For this purpose, we use a numerical reconstruction approach~(Figure~\ref{fig:figure04}). In general, the EQE as a function of wavelength can be written as
\begin{equation}
\text{EQE}(\lambda) = \int_0^d g(x,\lambda) f_c(x) \diffd x,
\label{eq:EQE}
\end{equation}
where $g(x,\lambda)$ is the local generation profile of free charges, which we estimate from the simulated absorption~$A(x,\lambda)$ times a constant factor summarizing all elementary steps prior to charge collection, that is, exciton diffusion, charge transfer and charge separation. Because Equation~(\ref{eq:EQE}) cannot be solved directly for~$f_c(x)$, we assume a simple step function with perfect collection~($f_c = 1$) at the anode side and zero collection at the cathode side. This shape is motivated by the simulated collection probability in Figure~\ref{fig:figure01}b, as well as previous works on doped organic blends\cite{Dibb2013} and quantum dot solids.\cite{Scheunemann2015}

Here, we advance these approaches by allowing the collection probability to become a function of the generation rate~(Figure~\ref{fig:figure04}a). For each light intensity, we let the position of the step vary until the best fit to the experimental EQE spectrum is found. Figure~\ref{fig:figure04}b shows the results of this analysis for the 300-nm thick device with opaque top electrode. Even though a step function is a simplification of the real collection probability, the reconstruction approach provides reasonably good fits over the whole intensity range and captures all important spectral features. In Figure~\ref{fig:figure04}c, we plot the fitted step position as a function of light intensity. Under 1-sun conditions, carriers are only collected from roughly half of the active layer, which reasonably explains the poor device performance. It should be noted that the effect is further enhanced by the device architecture; in the inverted devices studied herein, most of the incoming light is absorbed outside the collection zone.

\begin{table}
\caption{Simulation parameters for P3HT:PCBM.}
\begin{tabular}{lc}
\toprule
Parameter & Value\\
\midrule
Effective band gap [eV] & 1.1\\
Relative permittivity & 3.5\\
Effective density of states [$\unit{cm^{-3}}$] & $10^{20}$\\
Recombination coefficient, $k_2$ [$\unit{cm^{3}\,s^{-1}}$] & $2 \times 10^{-12}$\\
Electron mobility, $\mu_n$ [$\unit{cm^2\,V^{-1}\,s^{-1}}$] & $1.3 \times 10^{-3}$\\
Hole mobility, $\mu_p$ [$\unit{cm^2\,V^{-1}\,s^{-1}}$] & $1.1 \times 10^{-4}$\\
Injection barrier height [$\unit{eV}$] & $0.1$\\
\bottomrule
\end{tabular}
\label{tab:params}
\end{table}

Next, we are interested in whether the measured collection zone is correlated with the SCR due to imbalanced transport. For this purpose, we modeled the P3HT:PCBM device with the drift--diffusion model using the parameter set given in Table~\ref{tab:params}. The recombination coefficient and the carrier mobilities were experimentally determined, as detailed in the Supporting Information. To account for the non-trivial generation profile, we coupled the drift--diffusion simulator with the transfer-matrix model. As can be seen from Figure~\ref{fig:figure04}c, the numerically modeled collection zone~(open symbols) coincides well with the experimental one from the EQE reconstruction~(closed symbols). From this we can draw two important conclusions. First, with our simple and parameter-free approach based on EQE~measurements and a purely optical model, one can track the build-up of space charge in OPVs. Second, because the doping density was set to zero in the drift--diffusion simulation, the only possible source of space charge is the imbalanced transport. Hence, with our reconstruction approach, we are able to attribute the degraded device performance to a mobility mismatch without the need to know the values of the actual electron and hole mobility. Figure~\ref{fig:figure04}c also shows the analytical prediction according to Equation~(\ref{eq:SCR}) with $\mu_s = \mu_p = \unit[10^{-4}]{cm^2\,V^{-1}\,s^{-1}}$~(dashed line). The excellent agreement of experimental, numerical and analytical results further confirms that the collection zone from the EQE reconstruction can be assigned to the SCR caused by imbalanced charge transport in the active layer.

Another test to check if the device is limited by space charge is the thickness dependence of the photocurrent. If space charge effects are significant, $\jph$ should deviate from ideal behavior as soon as the thickness of the active layer exceeds the SCR width~($d > w$). In order to show that this is the case for the P3HT:PCBM devices, we fabricated a thickness series with~$d$ ranging from 65 to \unit[350]{nm}. As a figure of merit we use the scaling exponent~$\beta = (\diffd\jph/\diffd I)(I/\jph)$, which can be determined from the white-light bias EQE data with great precision~(see Experimental Section). As can be seen in Figure~\ref{fig:figure04}d, the photocurrent around \unit[1]{sun} intensity becomes deteriorated~($\beta < 1$) above a  thickness of about~\unit[150]{nm}, which coincides well with the calculated SCR width. Again, the experiment is well captured by the numerical drift--diffusion model. As indicated by the dashed line in Figure~\ref{fig:figure04}d, photocurrent losses would be less pronounced in a standard device architecture, where most carriers are generated within the SCR~(like in the `anode illumination' case in Figure~\ref{fig:figure03}). If we define the critical thickness as the point where $\beta$ drops below 0.975, it would be~$\approx\unit[150]{nm}$ for the inverted device and~$\approx\unit[175]{nm}$ for the standard device. Hence, for the given system with $\mu_p \ll \mu_n$, the standard architecture would allow for efficient charge collection in slightly thicker devices. Note that in none of the cases in Figure~\ref{fig:figure04}d, the scaling exponent saturates at the theoretical limit of $\beta = 3/4$. The deviation can be explained by the inhomogeneous generation profile, which was not considered in the derivation of Equation~(\ref{eq:jph}). This underlines once again that a test of the scaling behavior of the photocurrent alone is not sufficient to prove a device limitation by imbalanced transport.

\subsection{Thickness Limits due to Imbalanced Transport}
\begin{figure*}[t!]
\includegraphics[width=\linewidth]{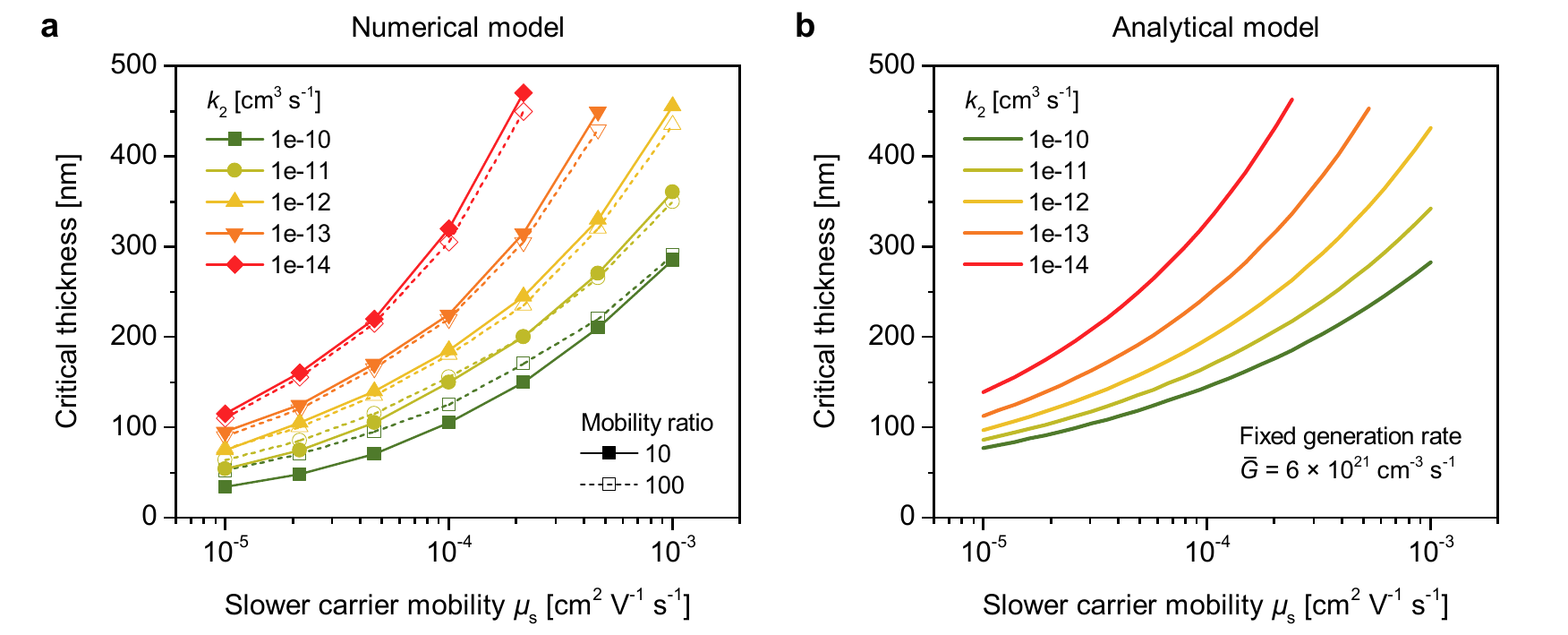}
\caption{Thickness limitations due to imbalanced charge transport. (a)~Numerically modeled thickness at which charge collection becomes degraded by space charge effects as a function of the slower carrier mobility~$\mu_s$ and for different recombination coefficients~$k_2$. Filled and open symbols correspond to a mobility ratio of 10 and 100, respectively. As operational definition for the critical thickness we take the point at which the scaling exponent $\beta$ drops below 0.975 around 1-sun illumination. The thickness dependence of the (spatially averaged) generation rate~$\bar{G}$ was synthesized from optical simulations for 12 different blend systems and 2 device architectures~(see the Supporting Information). (b)~Analytical model of the critical thickness~$\dcrit$ according to Equation~(\ref{eq:dcrit}) for a fixed generation rate of~$\bar{G} = \unit[6 \times 10^{21}]{cm^{-3}\,s^{-1}}$. The Debye screening length, determining the correction term~$\Delta w$, was approximated by $\LD = \sqrt{\varepsilon\varepsilon_0 kT/q^2 N_s}$, where $N_s = \sqrt{\varepsilon\varepsilon_0 G/q \mu_s}$.}
\label{fig:figure05}
\end{figure*}

Finally, we use our well calibrated numerical device model to discuss the general limits at which imbalanced transport will degrade the performance of OPV devices. For this purpose we determined the critical thickness as defined above for a range of mobilities and bimolecular recombination coefficients~(Figure~\ref{fig:figure05}a). This procedure requires one to make an assumption on how the generation rate varies with thickness. We therefore synthesized a function for the spatially averaged generation rate, $\bar{G} = (1/d)\int_0^d G(x)\diffd x$, versus the active-layer thickness~$d$ based on optical simulations for 12 different OPV blends, including both polymer\slash{}fullerene and polymer\slash{}non-fullerene systems, and 2 device architectures~(see the Supporting Information). Two important conclusions can be drawn from the simulations in Figure~\ref{fig:figure05}a. First, balanced transport is not a necessary condition for efficient thick-film OPVs if charge recombination is sufficiently reduced. For example, given a moderate mobility of $\unit[10^{-4}]{cm^2\,V^{-1}\,s^{-1}}$ for the slower carrier, a 300-nm thick solar cell that does not suffer from space charge effects would even be possible at a mobility imbalance of 10 or 100 if $k_2$ was about $\unit[10^{-13}]{cm^3\,s^{-1}}$. Such low recombination rates, which are orders of magnitude lower than what is predicted by the Langevin model, have already been demonstrated for OPV blends with a certain degree of molecular order.\cite{Heiber2018,Armin2016} Second, except in the case of very strong recombination~($k_2 \approx \unit[10^{-10}]{cm^3\,s^{-1}}$), the thickness limitation is independent on the actual mobility contrast and given by the mobility of the slower carrier only. This justifies the use of~$\mu_s$ instead of both the electron and hole mobility in Equations~(\ref{eq:jph}) and~(\ref{eq:SCR}) and is in line with earlier works highlighting the importance of the slower carrier mobility for charge transport and recombination.\cite{Mihailetchi2005,Kaienburg2016,Neher2016,Stolterfoht2015}

The reason why strongly non-Langevin systems can tolerate space charge effects to some extent is that some of the slower carriers will escape recombination within the quasi-neutral region and diffuse into the SCR where they are collected. Hence, the collection zone will be effectively extended by the diffusion length~$L_s$ of the slower carrier. Using the same arguments as in our recent paper on space charge induced by doping,\cite{Sandberg2019} the photocurrent can thus be written as~$\jph = -qG \times (w + L_s - \Delta w)$, where $w$ is the SCR width as given by Equation~(\ref{eq:SCR}) and $\Delta w$ a correction term that accounts for the recombination near the boundary between the SCR and the neutral region. It is therefore reasonable to analytically approximate the critical thickness as
\begin{equation}
\dcrit \approx w + L_s - \Delta w.
\label{eq:dcrit}
\end{equation}
In the scenario outlined in this work, in which only strictly bimolecular recombination is considered, the diffusion length is given by
\begin{equation}
L_s = \sqrt{\frac{\mu_s \tau k T}{q}},
\label{eq:Ls}
\end{equation}
where $kT$ is the thermal energy and $\tau  = (k_2 G)^{-1/2}$ the effective recombination lifetime. As further detailed in Ref.~\citenum{Sandberg2019}, the correction term~$\Delta w$ is mainly determined by the Debye screening length~$\LD$ and can be written as
\begin{equation}
\Delta w = \LD \sqrt{2 \ln\left(1 + \frac{\LD^2}{L_s^2}\right)}.
\label{eq:dW}
\end{equation}
We note that the correction is only significant for relatively strong recombination, while $\Delta w \rightarrow 0$ can be assumed for~$k_2 < \unit[10^{-11}]{cm^3\,s^{-1}}$. Figure~\ref{fig:figure05}b shows that although the thickness dependence of the generation rate is not explicitly taken into account, the analytical model in Equation~(\ref{eq:dcrit}) reproduces the trend of the numerical data reasonably well. This clearly demonstrates that diffusion of carriers from the field-free region can significantly contribute to the device photocurrent.

\section{Conclusions}
We have demonstrated and thoroughly tested a simple method to follow the build-up of space charge due to imbalanced transport in real operating OPVs. The only required input are EQE spectra and the optical constants, which are relatively straightforward to measure, while no information about the charge carrier mobilities and the recombination coefficient need not to be known. This makes the method suitable for screening of new OPV materials. Furthermore, we have provided numerical and analytical means to describe the effect of imbalanced charge transport in OPVs. Our theoretical framework shows that the paradigm that charge transport must be balanced can be overcome by sufficiently reducing charge recombination. Strongly non-Langevin systems can tolerate imbalanced transport even at thicknesses around \unit[300]{nm}. We therefore propose that reducing charge recombination is key for OPVs to become technologically relevant. This puts the questions of what factors influence the recombination and how it can be purposely suppressed among the most pressing ones regarding organic solar cells.

\section{Experimental Section}
\paragraph{Device Fabrication:} Solar cells were fabricated with the inverted device structure indium tin oxide~(ITO)\slash{}ZnO (\unit[40]{nm})\slash{}P3HT:PCBM (\unit[300]{nm})\slash{}top electrode. The ZnO layer consisted of nanoparticles with a diameter of \unit[5]{nm}, which were prepared as described elsewhere.\cite{Wilken2014}. Regioregular P3HT was purchased from Rieke~(4002-E) and PCBM from Solenne. Blend films of P3HT:PCBM were spin-coated from chlorobenzene~($\unit[30:30]{mg\,mL^{-1}}$) and thermally annealed at~\unit[150]{$^\circ$C} for \unit[10]{min}. The top electrode consisted of \ce{MoO3} (\unit[12]{nm})\slash{}Ag (\unit[150]{nm}) for the opaque devices and of \ce{MoO3} (\unit[12]{nm})\slash{}Au (\unit[12]{nm})\slash{}\ce{MoO3} (\unit[50]{nm}) for the semitransparent devices and was thermally evaporated under high vacuum~(\unit[$10^{-6}$]{mbar}). The device area was about~$\unit[0.3]{cm^3}$. For the thickness series, the concentration of the blend solution and the spin-coating speed were varied. All preparation steps were carried out in a nitrogen-filled glovebox. Devices were encapsulated with glass slides and an UV-curable adhesive prior to the measurements.

\paragraph{Measurements:} Current-voltage curves were measured with a Keithley~4200 parameter analyzer. Simulated AM1.5~illumination with an intensity of $\unit[100]{mW\,cm^{-2}}$ was provided by a class~AAA solar simulator~(Photo Emission Tech) and attenuated with neutral density filters. Carrier mobilities and the bimolecular recombination coefficient of the P3HT:PCBM blends were determined via the analysis of space-charge limited currents\cite{Felekidis2018} and charge extraction experiments,\cite{Scheunemann2017,Hoefler2018} as detailed in the Supporting Information. White-light bias EQE measurements were performed with a custom-built Bentham PVE300 system, equipped with a 75-W Xe arc lamp and a monochromator. Photocurrent signals were modulated at \unit[780]{Hz} and monitored with a lock-in amplifier~(Stanford Research SR830). Additional white-light bias illumination was provided by a 50-W halogen lamp. The intensity of the bias light was adjusted using a KG5-filtered, calibrated silicon solar cell~(Fraunhofer ISE). Because of the lock-in detection, the EQE experiment probes the modulation of the photocurrent, $\diffd\jph$, with respect to the modulated, monochromatic probe illumination~$\diffd I(\lambda)$. Hence, the quantity measured is the differential spectral responsivity, $\Delta S = \diffd \jph / \diffd I(\lambda)$, which deviates from the total spectral responsivity~($S = \jph/I$) when $\jph$ scales nonlinearly with the light intensity, $\jph \propto I^\beta$ with $\beta < 1$. Since $\Delta S$ and $S$ are approximately related by $\Delta S = \beta S$, the exponent $\beta$ can be derived from the ratio~$\Delta S/S$ at different (background) light intensities. While $\Delta S(\lambda,I)$ is directly measured, $S$ can be estimated by integrating $\Delta S$ with respect to~$I$, as described elsewhere.\cite{Cowan2013,Wehenkel2012} Hence, at given white-light bias intensity $I'$, the scaling exponent can be approximated by~$\beta(I') \cong \Delta S/(1/I')\int_{0}^{I'} \Delta S \diffd I$.

\paragraph{Drift--Diffusion Model:} One-dimensional drift--diffusion simulations were performed using the software SCAPS developed by Burgelman et al.\cite{Burgelman2000} The bulk heterojunction layer was approximated as an effective semiconductor placed between two Ohmic contacts, the electron-collecting cathode at~$x = 0$ and the hole-collecting anode at~$x = d$. Charge carrier injection was treated by thermionic emission. Charge carrier recombination was described by the empirical rate equation~$R = k_2(np - n_i^2)$, where $k_2$ is the bimolecular recombination rate constant and $n_i$ is the intrinsic carrier density. For the simulations in Figure~\ref{fig:figure01}, $k_2 = \unit[10^{-12}]{cm^{3}\,s^{-1}}$ and a constant generation of $\bar{G} = \unit[3 \times 10^{21}]{cm^{-3}\,s^{-1}}$ at 1-sun illumination was assumed. For the simulations of the P3HT:PCBM devices, the generation profile~$G(x)$ was calculated with the transfer-matrix model. 

\paragraph{Transfer-Matrix Model:} The absorption profiles~$A(\lambda,x)$ were simulated with a MATLAB program code based on the one-dimensional transfer-matrix method.\cite{Burkhard2010,Pettersson1999} The code was customized to account for the illumination direction and the position of the glass substrate in the semitransparent devices. The optical constants~(refractive index, extinction coefficient) of all materials involved were determined with a spectroscopic ellipsometer~(Woolam VWASE) as described previously.\cite{Scheunemann2015,Wilken2015}

\begin{acknowledgement}
We thank J\"urgen Parisi for his valuable contributions to the field of organic photovoltaics and his personal support of much of the research presented herein. We are also grateful to Marc Burgelman for assistance with the drift--diffusion simulations and to Martijn Kemerink for helpful discussion on the modelling of the charge-collection probability. This work received funding through the \AA{}bo Akademi University Research Mobility Programme and the Magnus Ehrnrooth Foundation.
\end{acknowledgement}

\section*{Keywords}
Organic photovoltaics, space charge, imbalanced charge transport, quantum efficiency, thick films

\bibliography{ms}

\end{document}


\newpage
\paragraph{Analytical model.} We consider the case with imbalanced carrier mobilities. In the case when holes are the slower carrier type, having a mobility $\mu_p = \mu_s$, a space charge region~(SCR) of photo-induced holes will be created in the active layer adjacent to the anode at high enough light intensities. This space charge region will screen the electric field in the remainder of the active layer where charge neutrality prevails. Under these conditions, the Poisson equation for the electric field~($F$) can be approximated by 
\begin{equation}
\frac{dF}{dx} = \frac{q}{\varepsilon\varepsilon_0} p(x)
\end{equation}
for $x < w < d$, where $w$ is the thickness of the SCR, $d$ is the thickness of the active layer, $x$ is the distance from the anode contact, $p$ is the hole density, $q$ is the elementary charge, and $\varepsilon\varepsilon_0$ is the permittivity of the active layer. The hole density is determined by the light intensity via the steady-state continuity equation for holes,
\begin{equation}
\frac{1}{q}\frac{dj_p}{dx} = G - R,
\end{equation}
where $j_p(x)$ is the hole current density, $G$ is the photogeneration rate of free carriers, and $R$ is the recombination rate. The corresponding electron current density is given by $j_n(x) = j-j_p(x)$, where $j$ is the total photo-induced current density which is independent of $x$. For simplicity, a uniform generation rate is assumed. Then, noting that the recombination of holes is negligible~($R=0$) within the space charge region~(since $p \gg n$), and taking the hole extraction to be dominated by drift~($j_p = q \mu_s p F$), the hole current can be expressed as 
\begin{equation}
j_p(x) = q \mu_s p(x) F(x) = qG[x-w]
\end{equation}
for $x<w$, assuming the charge collection of holes from the~(field-free) neutral region to be negligible. Upon eliminating the hole density in the Poisson equation, so that $dF/dx = qG[x-w]/(\mu_s\varepsilon\varepsilon_0 F)$, and integrating, the electric field inside the SCR is obtained as 
\begin{equation}
F(x) = \sqrt{\frac{qG}{\mu_s\varepsilon\varepsilon_0}} (x-w)
\end{equation}
assuming $F(x)=0$ for $w<x<d$. The electric field is then related to the applied voltage~($V$) via
\begin{equation}
V-V_0 = \int_0^w F dx
\end{equation}
where $V_0$ is the potential difference across the SCR at $V=0$. Hence, for the width of the SCR~region we obtain 
\begin{equation}
w = \sqrt{2(V_0-V)}\left(\frac{\varepsilon\varepsilon_0\mu_s}{qG}\right)^{1/4}.
\end{equation}
Finally, assuming the surface recombination~(of electrons) to be negligible, $j_n(0) = j-j_p(0) = 0$, the total photocurrent takes the form
\begin{equation}
j =-qG^{3/4} \left(\frac{4\mu_s\varepsilon\varepsilon_0}{q}\right)^{1/4} (V_0-V)^{1/2}.
\end{equation}
A completely analogous treatment applies for the case when electrons are the slower carrier type, resulting in identical expressions for $w$ and $j$, with $\mu_s = \min(\mu_n,\mu_p)$ being the mobility of the slower carrier type.

\newpage
\begin{figure*}
\includegraphics[width=\linewidth]{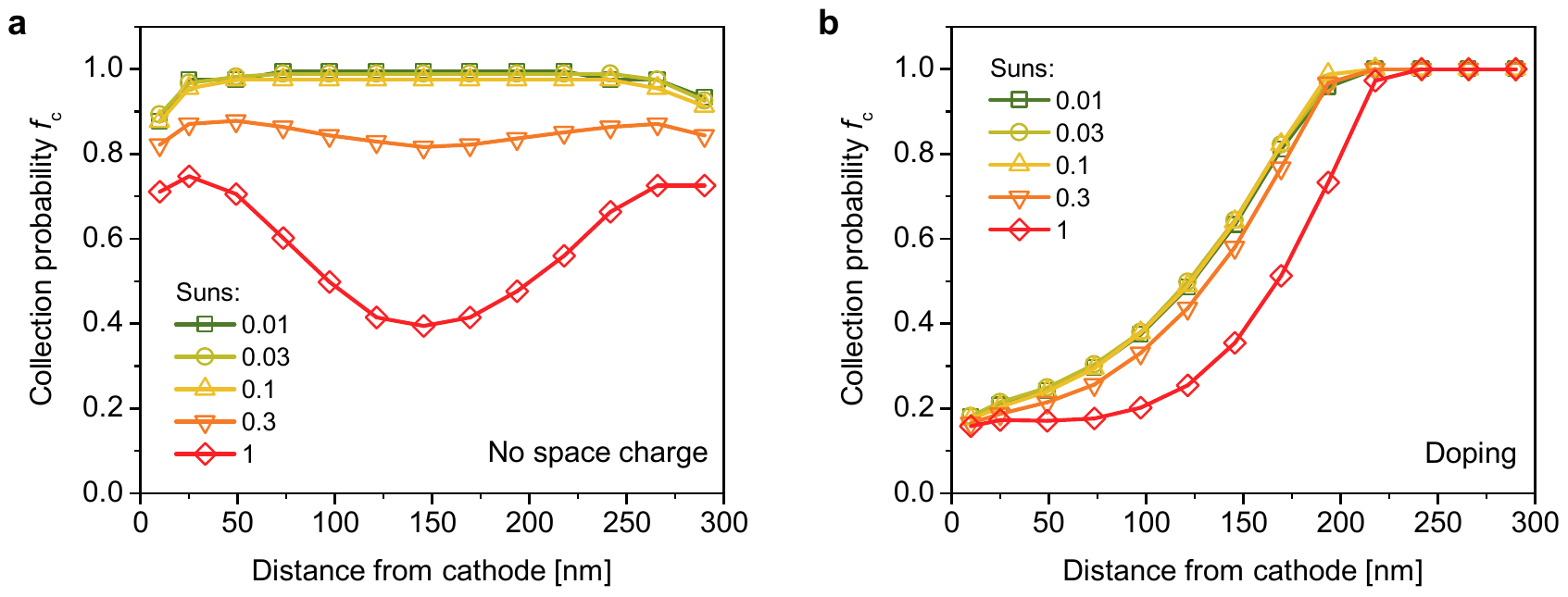}
\caption{Simulated collection probability for a thick-film device (a)~without space charge effects and (b)~with an n-doped active layer. In both cases, balanced carrier mobilities of $\mu_n = \mu_p = \unit[10^{-4}]{cm^2/Vs}$, a recombination coefficient of $k_2 = \unit[2 \times 10^{-12}]{cm^3/s}$, and a constant photogeneration rate of $G = \unit[3 \times 10^{21}]{cm^{-3}/s}$ were assumed. The doping concentration for the simulations in panel~b was $N_D = \unit[2\times10^{16}]{cm^{-3}}$.}
\label{fig:figureS1}
\end{figure*}

\newpage
\begin{figure*}
\includegraphics[width=\linewidth]{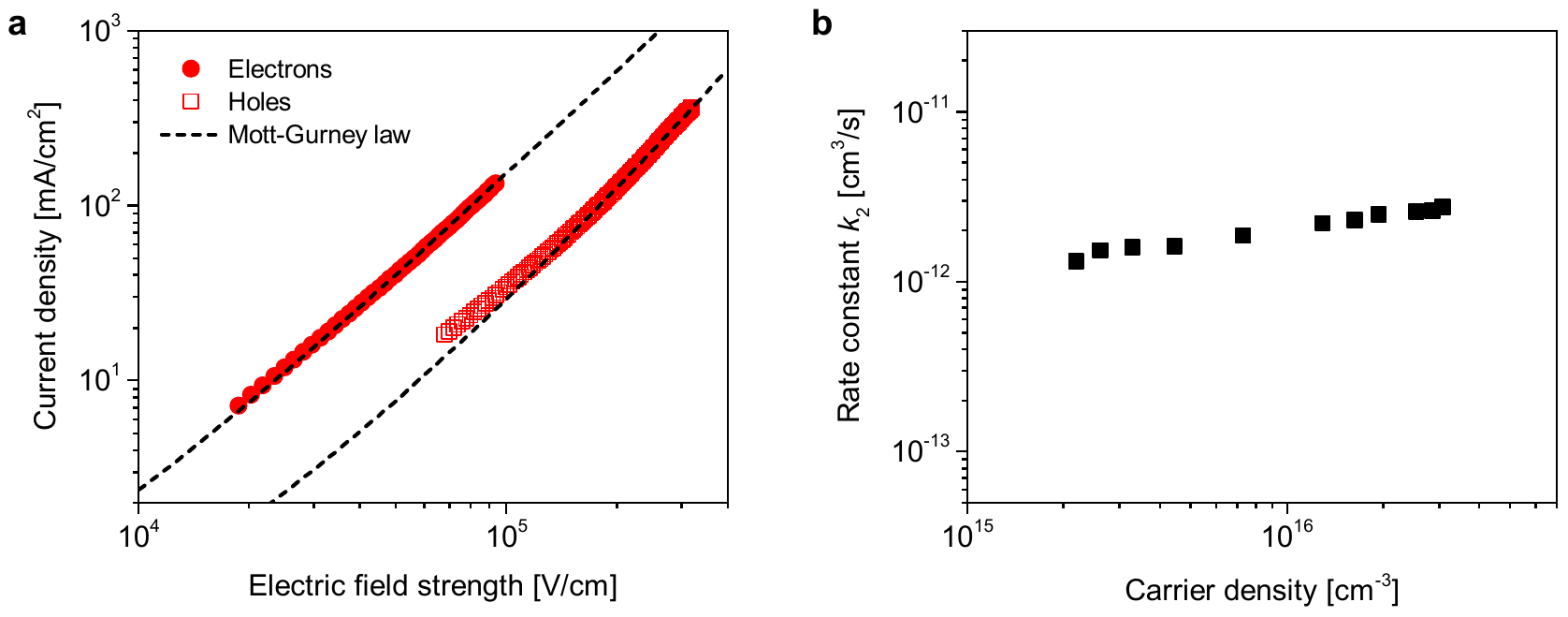}
\caption{Transport and recombination in P3HT:PCBM blends. (a)~Current--voltage curves of electron-only and hole-only devices with an active-layer thickness of~\unit[250]{nm}. Dashed lines are fits to the Mott--Gurney law, which give an electron mobility of $\mu_n = \unit[1.3 \times 10^{-3}]{cm^2/Vs}$ and a hole mobility of $\mu_p = \unit[1.1 \times 10^{-4}]{cm^2/Vs}$. Fitting was done with the software by Felekidis~et~al.\cite{Felekidis2018} (b)~Bimolecular recombination coefficient~$k_2$ as derived from combined bias-assisted charge extraction~(BACE) and transient photovoltage~(TPV) measurements. Experimental details of BACE and TPV are given in Refs.~\citenum{Scheunemann2017} and \citenum{Hoefler2018}.}
\label{fig:figureS2}
\end{figure*}

\newpage
\begin{figure*}
\includegraphics[width=\linewidth]{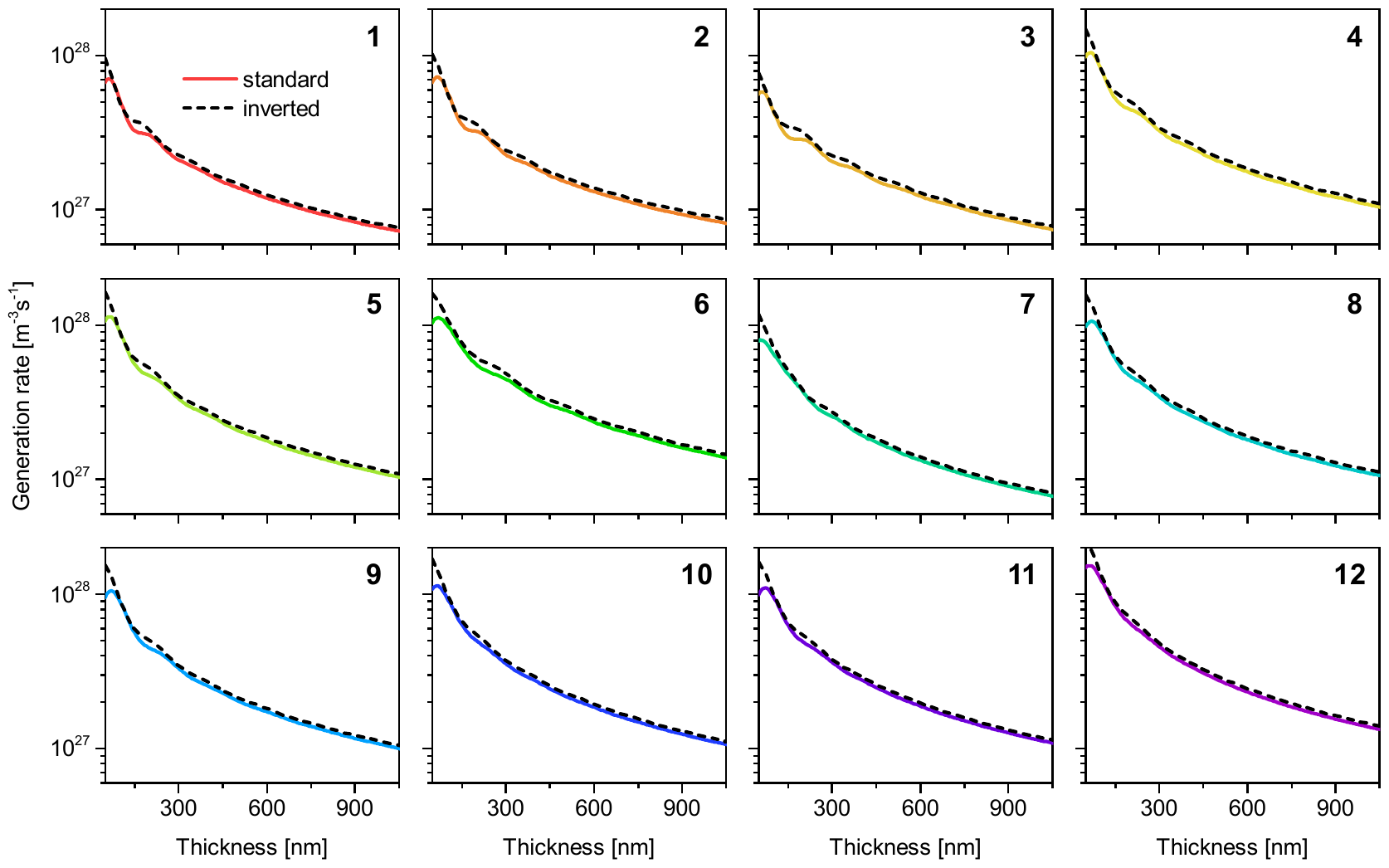}
\caption{Simulated thickness dependence of the average generation rate~$\bar{G}$ for standard and inverted devices of different OPV blends. (\textbf{1})~P3HT:PC61BM, this work. (\textbf{2})~P3HT:ICBA, Chen et~al.\cite{Chen2014} (\textbf{3})~PCDTBT:PC71BM, Sun et~al.\cite{Sun2011} (\textbf{4})~PTB7:PC71BM, Cho et~al.\cite{Cho2016} (\textbf{5})~PTB7-Th:PC71BM, Mescher et~al.\cite{Mescher2015} (\textbf{6})~LBG:PCBM, Chen et~al.\cite{Chen2014} (\textbf{7})~BTR:PC71BM, Armin et~al.\cite{Armin2016} (\textbf{8})~PTZ1:ITIC, Xiao et~al.\cite{Xiao2018} (\textbf{9})~PBDT(T)[2F]T:ITIC, Firdaus et~al.\cite{Firdaus2019} (\textbf{10})~PBDB-T:ITIC, Mao et~al.\cite{Mao2019} (\textbf{11})~PBDB-T-SF:IT-4F, Firdaus et~al.\cite{Firdaus2019} (\textbf{12})~PBDB-T:Y16, Ma et~al.\cite{Ma2019} The assumed device architectures were ITO\slash{}PEDOT:PSS\slash{}active layer\slash{}\ce{LiF}\slash{}Al~(standard) and ITO\slash{}ZnO\slash{}active layer\slash{}\ce{MoO3}\slash{}Ag~(inverted).}
\label{fig:figureS3}
\end{figure*}

\newpage
\begin{figure*}
\includegraphics[width=0.6\linewidth]{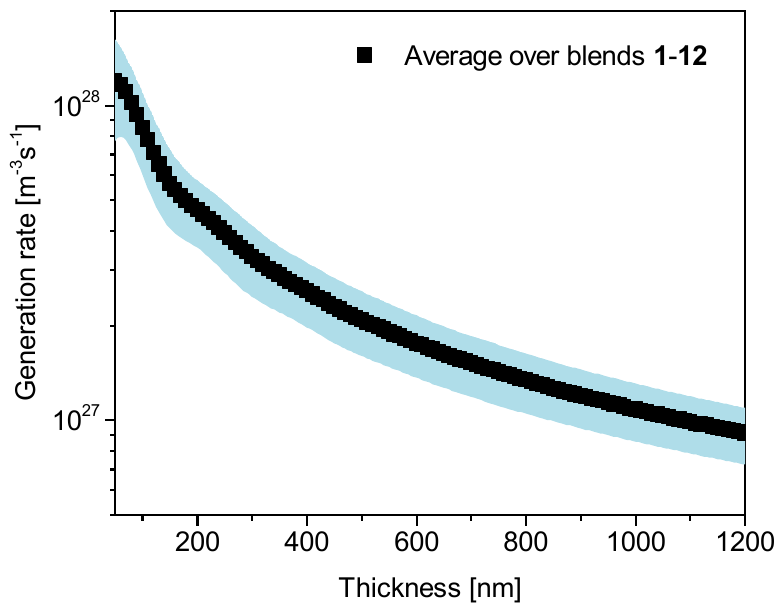}
\caption{Average over the generation rate versus thickness plots in Figure~\ref{fig:figureS3}. Shaded area is the standard deviation. This artificial generation rate is used as input for the simulations in Figure~5 in the main text.}
\label{fig:figureS4}
\end{figure*}

\bibliography{supplement}